\documentclass[%
 aps,
 pra,
 reprint,
 amsmath,
 amssymb,
 showkeys,
 groupedaddress,
 eprint,
 longbibliography,
 nofootinbib,
 nobibnotes,
]{revtex4-1}

\usepackage[utf8]{inputenc}
\usepackage{graphicx}
\usepackage{bm}
\usepackage{xcolor}
\usepackage{enumerate}
\usepackage{amsfonts, amsmath, amssymb}
\usepackage[english]{babel}
\usepackage{dsfont}
\usepackage{placeins}
\usepackage{comment}

\usepackage{hyperref}
\hypersetup{
	colorlinks = true,
	linkcolor = blue,
	citecolor = blue,
	urlcolor = blue,
	filecolor = black,
	linktoc = all,
}

\newcommand{\ent}{\textsc{qc}}
\newcommand{\qc}{\textsc{qc}}
\newcommand{\VN}{\textrm{VN}}
\newcommand{\dd}{\delta}

\newcommand{\HH}{\mathcal{H}}
\newcommand{\CC}{\mathcal{C}}

\newcommand{\one}{\mathds{1}}
\newcommand{\tr}{\textrm{tr}}
\newcommand{\ket}[1]{\left| #1 \right\rangle}
\newcommand{\bra}[1]{\left\langle #1 \right|}
\newcommand{\Phat}{P}

\newcommand{\ketbra}[1]{\ket{#1}\!\bra{#1}}

\newcommand{\G}{\rm G}
\newcommand{\QQ}{\mathcal{K}}

\begin{document}


\title{Basics of observational entropy}

\author{Joseph Schindler}
\email{jcschind@ucsc.edu}

\affiliation{SCIPP and Department of Physics, University of California, Santa Cruz, California 95064, USA}

\date{June 2020}

\begin{abstract}
These notes provide a brief primer on the basic aspects of ``observational entropy'' (also known as ``quantum coarse-grained entropy''), a general framework for applying the concept of coarse-graining to quantum systems. We review the basic formalism, survey applications to thermodynamics, make a connection to quantum correlations and entanglement entropy, compare to the corresponding classical theory, and discuss a generalization based on POVM measurements.
\end{abstract}


\maketitle



\section{Introduction}

Many related but distinct notions of entropy are important in physics. These range from informational measures of inherent uncertainty in a state, such as classical Gibbs and quantum von Neumann entropy, to ``microstate-counting entropies,'' such as classical Boltzmann entropy, and ultimately to thermodynamic entropy with applications to heat and work.

It is well known that the relationship between informational entropies and thermodynamic entropy is related to the concept of coarse-graining, as is the case with classical Boltzmann entropy. And indeed, examples of methods involving coarse-grained entropies can be found throughout the literature~(\textit{e.g.}~\cite{gibbs1902elementary,vonNeumann1948,wehrl1978general,penrose1979foundations,tolman1979principles,goldstein1981nonequilibrium,grabert1982projection,tabor1989chaos,gaspard1997entropy,gu1997time,falcioni2005production,gellmann2007quasiclassical,kozlov2007fine,saar2007fluctuation,shell2008relative,puglisi2010entropy,gemmer2014entropy,kelly2014coarse,alonso2017coarse,engelhardt2018decoding,busiello2019entropy} and many others). 

Recently, a precise formulation of coarse-graining applicable to general quantum systems, which was originally discussed by von Neumann~\cite{vonNeumann1948} (see~\cite{strasberg2020heat} for a more detailed history), has been given a thorough re-investigation~\cite{Safranek2019a,Safranek2019b} and shown to provide a comprehensive framework for connecting quantum entropies to thermodynamics~\cite{Safranek2019a,Safranek2019b,safranek2019classical,strasberg2020heat,strasberg2020dissipation,strasberg2019entropy,schindler2020entanglement}. This framework goes by the name ``observational entropy,'' or ``quantum coarse-grained entropy.'' Here we use the former name, to distinguish from other coarse-graining methods.

Among the many coarse-graining methods existing in the literature (see \textit{e.g.}~those cited above), many are equivalent to observational entropy. A key aspect of the present program is to explicitly identify observational entropy as thermodynamic entropy, in and out of equilibrium, and based on this to develop a unified description of quantum entropy, information, and thermodynamics. This broad framework therefore incorporates many existing results based on coarse-graining.

The motivation underlying the definition of observational entropy is to unify important aspects of various types of entropy. Toward this end, there has been progress in showing that observational entropy:
\begin{itemize}
    \item Defines thermodynamic entropy both in and out of equilibrium, such that the non-equilibrium entropy dynamically approaches the appropriate equilibrium value as a closed system thermalizes~\mbox{\cite{Safranek2019a,Safranek2019b,safranek2019classical}}, and also describes standard situations in thermodynamics such as a system coupled to a heat bath~\cite{strasberg2020heat,strasberg2020dissipation}.
    \item Generically increases, in the sense of the second law of thermodynamics~\cite{Safranek2019a,Safranek2019b,safranek2019classical,strasberg2020heat,strasberg2020dissipation}.
    \item Has a clear informational meaning, incorporating both a ``microstate counting'' (Boltzmann) contribution and a ``probabilistic uncertainty'' (Shannon) contribution~\cite{Safranek2019a,Safranek2019b}.
    \item Has a clear connection to the von Neumann~\cite{Safranek2019a,Safranek2019b} and entanglement~\cite{schindler2020entanglement} entropies.
    \item Has a direct correspondence with a classical version of the theory~\cite{safranek2019classical}.
\end{itemize}

Given these properties, this framework promises to provide an avenue toward a unified description of quantum entropy and thermodynamics. The goal of this paper is to provide a simple introduction to and primer on the subject.

\section{Basic formalism}

Observational entropy $S_\CC(\rho)$ assigns a value of entropy to a quantum state $\rho$ given a coarse-graining $\CC$. It represents an uncertainty of the state in the coarse-grained description. Different coarse-grainings are useful for describing different physical situations. The basic formalism is as follows.

A \textbf{coarse-graining} $\CC=\{\Phat_i\}$ is a collection of Hermitian ($\Phat_i^\dag = \Phat_i$) orthogonal projectors ($\Phat_i \Phat_j = \Phat_i \, \dd_{ij}$) forming a partition of unity ($\sum_i \Phat_i =  \one$). Each subspace generated by $\Phat_i$ is called a ``macrostate.'' 

One way to specify a coarse-graining is via the spectral decomposition of an observable operator $Q = \sum_q q \, \Phat_q$. The associated coarse-graining is $\CC_{Q} = \{ \Phat_q \}$, so that each macrostate corresponds to a different measurement outcome. Any coarse-graining can be associated with such an operator, so any coarse-graining can be seen as arising from measurement.

Given a coarse-graining $\CC$ the \textbf{observational entropy} of a density operator $\rho$ (on a Hilbert space $\HH$) is defined as
\begin{equation}
\label{eqn:obs-ent}
    S_{\CC}(\rho) = -\sum_i p_i \log \left( \frac{p_i}{V_i} \right),
\end{equation}
where $p_i = \tr(\Phat_i \rho)$ is the probability to find $\rho$ in each macrostate, and $V_i = \tr(\Phat_i)$ is the ``volume'' of each macrostate.%
\footnote{Equivalently one can define a coarse-grained density matrix $\rho_{\rm cg} = \sum_i p_i \Phat_i /V_i$ and let $S_{\CC}(\rho)= S^{\VN}(\rho_{\rm cg})$. The definition~(\ref{eqn:obs-ent}), however, is more consistent with the attitude adopted here that observational entropy and von Neumann entropy represent conceptually distinct types of uncertainty.
}

The definition can be rewritten in a form suggestive of its \textbf{informational meaning}
\begin{equation}
\label{eqn:suggestive}
    S_{\CC}(\rho) = -\sum_i p_i \log p_i + \sum_i p_i \log V_i.
\end{equation}
The first term is the Shannon entropy of the probability distribution over macrostates. The second term is the mean Boltzmann entropy.

There is a simple relation between observational entropy and von Neumann entropy: \textbf{von Neumann entropy} is the minimum value of observational entropy, minimized over coarse-grainings for a given $\rho$. Specifically, 
\begin{equation}
\label{eqn:vn-is-min}
    \inf_\CC \Big(S_{\CC}(\rho) \Big) = S_{\CC_{\rho}}(\rho) = S^{\VN}(\rho),
\end{equation}
where $S^{\VN}(\rho) = - \tr(\rho \log \rho)$ is the von Neumann entropy, and $\CC_\rho$ is the coarse-graining arising from spectral decomposition of $\rho$.  Thus (\ref{eqn:vn-is-min}) also expresses that no measurement is more informative than measuring the density matrix itself.

Coarse-grainings have a \textbf{partial ordering} where they may be ``finer,'' ``coarser,'' or incomparable to one another. We say $\CC = \{ P_i \}$ is coarser than $\CC' = \{ P_{i}^{'} \}$, written $\CC \geq \CC'$, if each $P_i \in \CC$ can be written as the sum $P_i = \sum_j Q_{j}^{'}$ of some subset $\{ Q_j \} \subset \CC'$. (That is, if each $P_i \in \CC$ can be partitioned into elements of $\CC'$.) If $\CC \geq \CC'$ then 
\begin{equation}
    S_{\CC}(\rho) \geq S_{\CC'}(\rho).
\end{equation}
In other words, refining a coarse-graining to a finer one does not increase observational entropy.

The coarsest coarse-graining is $\{ \one \}$, which yields $S_{\{\one\}}(\rho) = \log \dim \HH$. The finest coarse-grainings are those consisting of rank-1 projectors (that is, $V_i = 1$), corresponding to a measurement that fully distinguishes some orthonormal basis. In light of (\ref{eqn:vn-is-min}), $S_{C_\rho}(\rho)$ yields the minimum entropy of any coarse-graining. This statement can be strengthened by the additional statement $S_{\CC}(\rho) = S^{\VN}(\rho)$ if and only if $\CC_{\rho} \geq \CC$. These facts may be summarized by the \textbf{general bound}
\begin{equation}
    \label{eqn:bound}
   \log \dim \HH \geq  S_{\CC}(\rho) \geq S^{\VN}(\rho) ,
\end{equation}
where lower equality holds if and only if $\CC$ is finer than or equal to $\CC_\rho$ (the coarse-graining in eigenspaces of $\rho$).

In the same way that describing a system based on a measurement leads to a coarse-grained description, a system can be described by a sequence of measurements, leading to a description by ``multi-coarse-graining.'' A \textbf{multi-coarse-graining} $\vec{\CC}=(\CC^1, \CC^2, \ldots, \CC^3)$ is simply a sequence of coarse-grainings.

The \textbf{observational entropy in a multi-coarse-graining} $\vec{\CC}=(\CC_1,\CC_2)$ (the case of two coarse-grainings generalizes immediately to longer sequences), where $\CC_1=\{\Phat_{i}^{1}\}$ and $\CC_2=\{\Phat_{j}^{2}\}$, is defined by
\begin{equation}
\label{eqn:obs-ent-joint}
    S_{\vec{\CC}} \, (\rho) = -\sum_{ij} p_{ij} \log \left( \frac{p_{ij}}{V_{ij}} \right),
\end{equation}
where $p_{ij} = \tr(\Phat_{j}^{2} \, \Phat_{i}^{1} \, \rho \, \Phat_{i}^{1} \, \Phat_{j}^{2} )$ and $V_{ij} = \tr(\Phat_{j}^{2}  \Phat_{i}^{1} \Phat_{i}^{1}  \Phat_{j}^{2} )$. Coarse-grainings are said to commute if all their projectors commute. Both commuting and non-commuting multi-coarse-grainings can be physically relevant. Order of the sequence is relevant in the non-commuting case. This formula obeys the same general bound as for a single coarse-graining.

A commuting multi-coarse-graining can be redefined as a single coarse-graining, called a \textbf{joint coarse-graining}, since in the commuting case the set of products $\{\Phat_{i}^{1} \, \Phat_{j}^{2} \}$ itself forms a coarse-graining.

One important class of coarse-grainings are the ``local coarse-grainings,'' defined when the Hilbert space can be decomposed to a tensor product $\HH = \HH_A \otimes \HH_B \otimes \ldots \otimes \HH_C$. \textbf{Local coarse-grainings} are defined by 
\begin{equation}
\label{eqn:local-cg}
    \CC_A \otimes \CC_B \otimes \ldots \otimes \CC_C = \{ \Phat^A_l \otimes \Phat^B_m \otimes \ldots \otimes \Phat^C_n \},
\end{equation}
where 
$\CC_A= \{ \Phat^A_l \}$
is a coarse-graining of $A$, and so on for the other subsystems. These are precisely the coarse-grainings describing subsystem-local measurements (\textit{i.e.} consisting of only local operators). Local coarse-grainings may, but need not, be viewed as commuting multi-coarse-grainings, since $(\overline{\CC}_A, \ldots, \overline{\CC}_C)$, where $\overline{\CC}_A = \{ P^A_l \otimes \one_{B \ldots C} \}$ and so on, yields the same probabilities.

An important property of local coarse-grainings is the identity
\begin{equation}
\label{eqn:product-fomula}
    S_{\CC_A \otimes \ldots \otimes \CC_C}(\rho) = 
    \Big( \sum_{X} S_{C_X}(\rho_X) \Big)
    - I_{\CC_A \otimes \ldots \otimes \CC_C}(\rho),
\end{equation}
where $X \in \{A,B,\ldots,C \}$ labels the subsystems, with $\rho_X$ the reduced density in each one. The first term is the sum of \textbf{marginal entropies}, and the second term
\begin{equation}
    I_{\CC_A \otimes \ldots \otimes \CC_C}(\rho) \equiv \sum_{lm \ldots n} p_{lm \ldots n} \log \left( \frac{p_{lm \ldots n}}{p^A_l p^B_m \ldots p^C_n} \right)
\end{equation}
is the \textbf{mutual information} of the joint measurement. The~$p^A_l \equiv \sum_{m\dots n}p_{lm\dots n}=\tr(\Phat^A_l\rho_A)$ and so on are marginal probabilities, and subadditivity of Shannon entropy implies $I \geq 0$.

This survey of the formalism of observational entropy is sufficient to understand most results within the framework. For further development of the general theory, and proofs of the identities given above, see~\cite{Safranek2019a,Safranek2019b}.

\section{Application to thermodynamics}

Observational entropy has so far been applied (focusing on studies explicitly formulated within the present framework) to study thermodynamics in several ways. Some basic results include:
\begin{itemize}
    \item Equilibrium thermodynamic entropy can be identified as observational entropy in an energy coarse-graining---that is, in the coarse-graining $\CC_H$ of eigenspaces of the Hamiltonian $H$. This is easily verified in both the microcanonical ($\rho \propto \sum_E \Phat_E$ is a sum of nearby energy projectors) and canonical ($\rho \propto e^{-\beta H}$) ensembles---since in these cases $\rho$ is simply a mixture of energy eigenspaces so that $S_{C_H}(\rho) = S^{\VN}(\rho)$, leading to the usual microcanonical and canonical entropies. Grand canonical entropy can be treated similarly, by jointly coarse-graining with additional conserved quantities.

    \item For an isolated system with local interactions, non-equilibrium thermodynamic entropy can be identified with observational entropy $S_{\CC}(\rho)$ in a \textit{local} energy coarse-graining $\CC = \otimes_i \, \CC_{H_i}$, where the system is split into small but macroscopic local subsystems each with local Hamiltonian $H_i$. This has been called the ``factorized observational entropy'' (FOE). Starting out of equilibrium, over time this entropy dynamically approaches the expected equilibrium value (up to some corrections dependent upon finite-size effects and on the initial state). Not only does FOE dynamically equilibrate, it also has a clear interpretation when the system has only partially equilibrated: at some intermediate time~$t$ it represents the equilibrium thermodynamic entropy the system would eventually attain if (hypothetically) at time $t$ all interaction between the local subsystems was turned off. This entropy therefore helps understand the thermalization of closed systems. See~\cite{Safranek2019a,Safranek2019b,safranek2019classical} for further discussion.
    
    \item For a system coupled to a heat bath, the total non-equilibrium thermodynamic entropy can be identified with the observational entropy $S_{\CC_S \otimes \CC_{E}}$,where $\CC_{E}$ is an energy coarse-graining of the bath, and $\CC_S$ is any coarse-graining of the system. This entropy was shown to obey standard thermodynamic laws for entropy production in open systems, and properly describe thermodynamics of the system-bath interaction. See~\cite{strasberg2020heat,strasberg2020dissipation} for further discussion.
    
    \item Observational entropy generically obeys a second law of thermodynamics in closed systems. That is, for generic coarse-grainings, observational entropy in a closed system typically increases to, then for long periods of time stays near, a maximum value. This is true for any coarse-graining not specifically chosen to give a constant observational entropy (for instance $\CC_Q$ where $[Q,H]=0$ gives a constant entropy). See~\cite{Safranek2019a,Safranek2019b,safranek2019classical} for further discussion.
    
    \item Thermodynamic entropy defined based on observational entropy obeys an appropriate second law of thermodynamics in the sense of heat transfer between an open system and a thermal bath. That is, total entropy is non-decreasing (and related to irreversiblity) during thermodynamic processes. It can also be related to the first law. See~\cite{strasberg2020heat,strasberg2020dissipation} for further discussion.
\end{itemize}

Expanding these results to a comprehensive treatment of thermodynamics with observational entropy continues as an active area of research.

\section{quantum correlations\\ and entanglement entropy}

In the context of observational entropy, entanglement entropy has a clear informational and statistical mechanical meaning not only for bipartite pure states, where it is usually defined, but also for general states in multipartite systems. This connection is furnished by a generalization of entanglement entropy called ``quantum correlation entropy'' (or ``quarrelation entropy'' for short). 

In a multipartite system whose Hilbert space decomposes to $\HH = \HH_A \otimes \HH_B \otimes \ldots \otimes \HH_C$, the \textbf{quantum correlation entropy} is defined by
\begin{equation}
\label{eqn:ee}
    S^{\ent}_{AB \ldots C}(\rho) = 
    \inf_{\CC = \CC_A \otimes \ldots \otimes \CC_C} \Big(S_{\CC}(\rho) \Big) - S^{\VN}(\rho),
\end{equation}
where the infimum is over all \emph{local} coarse-grainings (\textit{cf.}~(\ref{eqn:vn-is-min}) where the infimum is over \emph{all} coarse-grainings), and the subscript denotes a partition into subsystems, allowing various partitions of the same system.

This quantity represents the difference in observational entropy between the most informative local coarse-graining, as compared to the most informative global coarse-graining. It measures how much information is inaccessible to observers restricted to make subsystem-local measurements.

In the special case of a bipartite system $AB$ in a pure state $\rho = \ketbra{\psi_{AB}}$, $S^{\ent}$ reduces to the standard entanglement entropy
\begin{equation}
\label{eqn:ee-bipartite-pure}
    S^{\ent}_{AB}(\ketbra{\psi_{AB}}) = S^{\VN}(\rho_A) = S^{\VN}(\rho_B),
\end{equation}
where $\rho_A$, $\rho_B$ are the reduced densities in each subsystem. But in general the quarrelation entropy and subsystem von Neumann entropies are not equivalent.

The bipartite pure case shows that $S^{\ent}$ generalizes entanglement entropy. In mixed and multipartite states, however, $S^\ent$ measures more general quantum correlations. In particular it is zero on strictly classically correlated states (those that can be diagonalized in a locally orthonormal product basis), but can be nonzero on separable states. This clarifies that $S^\qc$ is a measure of total nonclassical correlations (similar to quantum discord~\cite{modi2010unified}), and not an entanglement measure.

The interpretation of quarrelation entropy in terms of observational entropy is furnished by the inequality
\begin{equation}
\label{eqn:ee-ineq}
    S_{\CC_A \otimes \ldots \otimes \CC_C}(\rho)\geq S^{\VN}(\rho) + S^{\ent}_{AB \ldots C}(\rho),
\end{equation}
which holds for any local coarse-graining $\CC_A \otimes \ldots \otimes \CC_C$. That is, any local description of a system includes contributions to the total entropy (of the joint system) both due to mixedness of $\rho$ (the von Neumann entropy) and due to nonlocal correlations between the subsystems (the entanglement entropy).

In the earlier discussion of thermodynamics, non-equilibrium thermodynamic entropy was given in both examples by observational entropy in a local coarse-graining. Thus one finds that $S^{\ent}$ is typically a contribution to non-equilibrium thermodynamic entropy.

The quantity $S^{\ent}$ is equal to another well known measure of non-classicality, the ``relative entropy of quantumness''~\cite{groisman2007quantumness,piani2011all}, which measures the relative entropy distance of a state from the nearest classically correlated state. These are also equal to still another thermodynamically motivated correlation measure, the ``zero-way quantum deficit''~\cite{horodecki2005local}, which relates to information/work extractable by local operations. 

The connection with these other measures shows that $S^{\ent}$ is a broadly useful quantity, applicable both within and beyond the present framework, which is important both as an entropy and as a measure of the non-classicality of a state.

See~\cite{schindler2020entanglement} for further discussion.

\section{Classical observational entropy}

Observational entropy can also be defined for classical systems, described by a phase space $\Gamma$, with generalized coordinates $q_k$ and canonical momenta $\pi_k$. The state is given by a classical probability density $\rho(\vec{q},\vec{\pi})$ on phase space. The classical form of observational entropy is basically a standard application of coarse-graining to classical systems---the generic increase of such an entropy is mentioned for instance in the review by Wehrl~\cite{wehrl1978general}.

In the classical case a coarse-graining is simply a partition of the phase space into disjoint subsets $\Gamma_i \subset \Gamma$. That is $\CC = \{ \Gamma_i \}$ such that $\bigcup_i \Gamma_i = \Gamma$ and, for $i \neq j$, $\Gamma_i \cap \Gamma_j = \emptyset$. Each subset $\Gamma_i$ is called a macrostate.

The definition (\ref{eqn:obs-ent}) of observational entropy is unchanged, but in the classical case the probabilities and volumes are defined as
\begin{equation}
    \label{eqn:classical-p-v}
    p_i = \int_{\Gamma_i} \rho \; d\vec{q} \, d\vec{\pi},
    \qquad
    V_i = \int_{\Gamma_i}  d\vec{q} \, d\vec{\pi}.
\end{equation}
A normalizing constant may also be included in the integration measure for consistency of units. The interpretation of classical and quantum observational entropies are essentially the same.

In the classical formalism the Gibbs entropy \mbox{$S^{\G}(\rho) = -\int_{\Gamma} \rho \log \rho \, d\vec{q} \, d\vec{\pi}$} replaces von Neumann entropy as the minimal observational entropy. This is a satisfying connection since each represents the essential uncertainty in the state itself.

Moreover, in the classical case there is a unique ``finest'' coarse-graining, in which every point of phase space (microstate) is considered to be its own macrostate. Observational entropy in this ``fine-graining'' yields the Gibbs entropy. (There are also coarser coarse-grainings---the coarsest being a classical version of $\CC_\rho$---that also yield the Gibbs entropy.)

Classical thermodynamics based on the classical coarse-grained entropy has been studied, and many of the same thermodynamic properties discussed above for the quantum case also hold in classical systems. In particular the connections to equilibrium thermodynamic entropy and the application to non-equilibrium thermodynamics of an isolated system through (a classical version of) FOE, as well as the second law of thermodynamics in closed systems, were all shown to hold in the classical case~\cite{safranek2019classical}.

Local coarse-grainings can also be defined in the classical case. The total phase space is viewed as a Cartesian product $\Gamma = \Gamma_1 \times \Gamma_2$ (the case of two subsystems directly generalizes to many), where $\Gamma_1$ has some number of the coordinates, say $q_i^1$, along with their canonical momenta $\pi^1_i$. Then ``local'' coarse-grainings arise as Cartesian products of subsystem coarse-grainings. However, by the definition (\ref{eqn:ee}), one finds that in the classical case entanglement entropy is always zero---this must be true since the ``finest'' classical coarse-graining discussed above is itself local. Thus when quantum correlation entropy contributes to non-equilibrium thermodynamic entropy, it is a purely quantum effect.

The close parallel between the classical and quantum formalisms helps illuminate the fundamental aspects of coarse-grained entropy, and also demonstrates where purely quantum effects, like entanglement/quarrelation entropy, become important.

See~\cite{safranek2019classical} for further discussion.


\section{Extension to POVM coarse-grainings}

By the above definitions, a coarse-graining is essentially a projective measurement. This can be generalized to coarse-grainings representing positive operator valued measurements (POVMs) as follows.

A \textbf{POVM coarse-graining} is a set $\QQ=\{K_i\}$ of trace-preserving (\mbox{$\sum_i K_i^\dag K_i = 1$}) operators (acting as Kraus operators). Then observational entropy becomes
\begin{equation}
    S_{\QQ}(\rho) = -\sum_i p_i \log \left( \frac{p_i}{V_i} \right),
\end{equation}
where
\begin{equation}
p_i = \tr(K_i \, \rho \, K_i^\dag),
\qquad
V_i = \tr(K_i K_i^\dag),
\end{equation}
are the ``likelihoods'' and ``volumes'' of measurement outcomes. In this case ``macrostate'' may be interpreted to mean ``measured state.'' Projective coarse-grainings arise in the special case $K_i^\dag = K_i$ and $K_i K_j = K_i \, \dd_{ij}$ (\textit{i.e.}~when the Kraus operators are Hermitian orthogonal projectors).

Some basic facts are known about POVM coarse-grainings, for example that $S_\QQ(\rho) \geq S_\VN(\rho)$ continues to hold, and that multi-coarse-grainings can be defined with reasonable properties~\cite{safranek2020observational}. But no comprehensive theory of POVM coarse-grainings has been established yet. Doing so remains an important direction of future research.

The extension to POVM coarse-grainings is particularly interesting in relation to the quantum correlation entropy $S^\qc$ discussed earlier. 

Local POVM coarse-grainings may be defined by
\begin{equation}
    \QQ_A \otimes \QQ_B \otimes \ldots \otimes \QQ_C = \{ K^A_l \otimes K^B_m \otimes \ldots \otimes K^C_n \},
\end{equation}
where 
$\QQ_A= \{ K^A_l \}$
is a POVM coarse-graining of $A$, and so on. The elements $K_{lm \ldots n}$ of the coarse-graining implement separable operations. Not all separable operations are included, however, since the operations must be locally as well as globally trace-preserving (that is, since it is required that $\sum_l (K^A_l)^\dag K^A_l = 1$ at each subsystem, in addition to $\sum_{lm \ldots n}K_{lm \ldots n}^\dag K_{lm \ldots n} =1 $).

By infimizing (\ref{eqn:ee}) over local POVM (rather than local projective) coarse-grainings one obtains a generalization of the quantum correlation entropy. An interesting open question is whether this generalization is equivalent to the projective case (that is, whether the most informative local coarse-graining is always projective, as is the case for global coarse-grainings). If not, it will be interesting to see what kind of correlations (\textit{e.g.} discord-type or entanglement-type) the generalized $S^\qc$ measures.

A classical analog of POVM coarse-grainings can also be defined. One simply replaces $\rho$ with the phase space density distribution (as earlier) and the operators $K_i$ with a set of complex functions $\chi_i$ on phase space such that $\sum_i |\chi_i|^2 = 1$ (also, traces become integrals over phase space). This represents classical measurements with more realistic uncertainty than the projective case.

\section{Conclusion}

The theory of coarse-graining and observational entropy, in both classical and quantum systems, provides a useful foundation for a comprehensive and unified theory of statistical mechanics in and out of equilibrium, which is still being actively developed. The goal of the present paper has been to provide a basic introduction to, and survey of, this subject.


\begin{acknowledgments}
Thanks to Dominik \v{S}afr\'{a}nek for introducing me to this field, and to DS again along with Anthony Aguirre and Josh Deutsch for their close collaboration. This research was supported by the Foundational Questions Institute and by the Faggin Presidential Chair Fund.
\end{acknowledgments}





\bibliography{basics}

\begin{thebibliography}{33}%
\makeatletter
\providecommand \@ifxundefined [1]{%
 \@ifx{#1\undefined}
}%
\providecommand \@ifnum [1]{%
 \ifnum #1\expandafter \@firstoftwo
 \else \expandafter \@secondoftwo
 \fi
}%
\providecommand \@ifx [1]{%
 \ifx #1\expandafter \@firstoftwo
 \else \expandafter \@secondoftwo
 \fi
}%
\providecommand \natexlab [1]{#1}%
\providecommand \enquote  [1]{``#1''}%
\providecommand \bibnamefont  [1]{#1}%
\providecommand \bibfnamefont [1]{#1}%
\providecommand \citenamefont [1]{#1}%
\providecommand \href@noop [0]{\@secondoftwo}%
\providecommand \href [0]{\begingroup \@sanitize@url \@href}%
\providecommand \@href[1]{\@@startlink{#1}\@@href}%
\providecommand \@@href[1]{\endgroup#1\@@endlink}%
\providecommand \@sanitize@url [0]{\catcode `\\12\catcode `\$12\catcode
  `\&12\catcode `\#12\catcode `\^12\catcode `\_12\catcode `\%12\relax}%
\providecommand \@@startlink[1]{}%
\providecommand \@@endlink[0]{}%
\providecommand \url  [0]{\begingroup\@sanitize@url \@url }%
\providecommand \@url [1]{\endgroup\@href {#1}{\urlprefix }}%
\providecommand \urlprefix  [0]{URL }%
\providecommand \Eprint [0]{\href }%
\providecommand \doibase [0]{https://doi.org/}%
\providecommand \selectlanguage [0]{\@gobble}%
\providecommand \bibinfo  [0]{\@secondoftwo}%
\providecommand \bibfield  [0]{\@secondoftwo}%
\providecommand \translation [1]{[#1]}%
\providecommand \BibitemOpen [0]{}%
\providecommand \bibitemStop [0]{}%
\providecommand \bibitemNoStop [0]{.\EOS\space}%
\providecommand \EOS [0]{\spacefactor3000\relax}%
\providecommand \BibitemShut  [1]{\csname bibitem#1\endcsname}%
\let\auto@bib@innerbib\@empty
\bibitem [{\citenamefont {Gibbs}(2010)}]{gibbs1902elementary}%
  \BibitemOpen
  \bibfield  {author} {\bibinfo {author} {\bibfnamefont {J.~W.}\ \bibnamefont
  {Gibbs}},\ }\href {https://doi.org/10.1017/CBO9780511686948} {\emph {\bibinfo
  {title} {Elementary Principles in Statistical Mechanics}}},\ Cambridge
  Library Collection - Mathematics\ (\bibinfo  {publisher} {Cambridge
  University Press},\ \bibinfo {year} {[1902] 2010})\BibitemShut {NoStop}%
\bibitem [{\citenamefont {{von Neumann}}(2010)}]{vonNeumann1948}%
  \BibitemOpen
  \bibfield  {author} {\bibinfo {author} {\bibfnamefont {J.}~\bibnamefont {{von
  Neumann}}},\ }\bibfield  {title} {\bibinfo {title} {{Proof of the ergodic
  theorem and the H-theorem in quantum mechanics. Translation of: Beweis des
  Ergodensatzes und des H-Theorems in der neuen Mechanik}},\ }\href
  {https://doi.org/10.1140/epjh/e2010-00008-5} {\bibfield  {journal} {\bibinfo
  {journal} {Eur. Phys. J. H}\ }\textbf {\bibinfo {volume} {35}},\ \bibinfo
  {pages} {201} (\bibinfo {year} {2010})},\ \Eprint
  {https://arxiv.org/abs/1003.2133} {arXiv:1003.2133 [physics.hist-ph]}
  \BibitemShut {NoStop}%
\bibitem [{\citenamefont {{Wehrl}}(1978)}]{wehrl1978general}%
  \BibitemOpen
  \bibfield  {author} {\bibinfo {author} {\bibfnamefont {A.}~\bibnamefont
  {{Wehrl}}},\ }\bibfield  {title} {\bibinfo {title} {{General properties of
  entropy}},\ }\href {https://doi.org/10.1103/RevModPhys.50.221} {\bibfield
  {journal} {\bibinfo  {journal} {Rev. Mod. Phys.}\ }\textbf {\bibinfo {volume}
  {50}},\ \bibinfo {pages} {221} (\bibinfo {year} {1978})}\BibitemShut
  {NoStop}%
\bibitem [{\citenamefont {Penrose}(1979)}]{penrose1979foundations}%
  \BibitemOpen
  \bibfield  {author} {\bibinfo {author} {\bibfnamefont {O.}~\bibnamefont
  {Penrose}},\ }\bibfield  {title} {\bibinfo {title} {Foundations of
  statistical mechanics},\ }\href {https://doi.org/10.1088/0034-4885/42/12/002}
  {\bibfield  {journal} {\bibinfo  {journal} {Rep. Prog. Phys.}\ }\textbf
  {\bibinfo {volume} {42}},\ \bibinfo {pages} {1937} (\bibinfo {year}
  {1979})}\BibitemShut {NoStop}%
\bibitem [{\citenamefont {Tolman}(1979)}]{tolman1979principles}%
  \BibitemOpen
  \bibfield  {author} {\bibinfo {author} {\bibfnamefont {R.~C.}\ \bibnamefont
  {Tolman}},\ }\href {http://cds.cern.ch/record/1546394} {\emph {\bibinfo
  {title} {{The principles of statistical mechanics}}}}\ (\bibinfo  {publisher}
  {Dover},\ \bibinfo {address} {New York, NY},\ \bibinfo {year}
  {1979})\BibitemShut {NoStop}%
\bibitem [{\citenamefont {{Goldstein}}\ and\ \citenamefont
  {{Penrose}}(1981)}]{goldstein1981nonequilibrium}%
  \BibitemOpen
  \bibfield  {author} {\bibinfo {author} {\bibfnamefont {S.}~\bibnamefont
  {{Goldstein}}}\ and\ \bibinfo {author} {\bibfnamefont {O.}~\bibnamefont
  {{Penrose}}},\ }\bibfield  {title} {\bibinfo {title} {{A nonequilibrium
  entropy for dynamical systems}},\ }\href {https://doi.org/10.1007/BF01013304}
  {\bibfield  {journal} {\bibinfo  {journal} {J. Stat. Phys.}\ }\textbf
  {\bibinfo {volume} {24}},\ \bibinfo {pages} {325} (\bibinfo {year}
  {1981})}\BibitemShut {NoStop}%
\bibitem [{\citenamefont {{Grabert}}(1982)}]{grabert1982projection}%
  \BibitemOpen
  \bibfield  {author} {\bibinfo {author} {\bibfnamefont {H.}~\bibnamefont
  {{Grabert}}},\ }\href {https://doi.org/10.1007/BFb0044591} {\emph {\bibinfo
  {title} {{Projection Operator Techniques in Nonequilibrium Statistical
  Mechanics}}}}\ (\bibinfo  {publisher} {Springer},\ \bibinfo {address}
  {Berlin, Heidelberg},\ \bibinfo {year} {1982})\BibitemShut {NoStop}%
\bibitem [{\citenamefont {{Tabor}}(1989)}]{tabor1989chaos}%
  \BibitemOpen
  \bibfield  {author} {\bibinfo {author} {\bibfnamefont {M.}~\bibnamefont
  {{Tabor}}},\ }\href@noop {} {\emph {\bibinfo {title} {Chaos and Integrability
  in Nonlinear Dynamics: An Introduction}}}\ (\bibinfo  {publisher}
  {WileyInterscience},\ \bibinfo {year} {1989})\BibitemShut {NoStop}%
\bibitem [{\citenamefont {{Gaspard}}(1997)}]{gaspard1997entropy}%
  \BibitemOpen
  \bibfield  {author} {\bibinfo {author} {\bibfnamefont {P.}~\bibnamefont
  {{Gaspard}}},\ }\bibfield  {title} {\bibinfo {title} {{Entropy production in
  open volume-preserving systems}},\ }\href
  {https://doi.org/10.1007/BF02732432} {\bibfield  {journal} {\bibinfo
  {journal} {J. Stat. Phys.}\ }\textbf {\bibinfo {volume} {88}},\ \bibinfo
  {pages} {1215} (\bibinfo {year} {1997})}\BibitemShut {NoStop}%
\bibitem [{\citenamefont {{Gu}}\ and\ \citenamefont
  {{Wang}}(1997)}]{gu1997time}%
  \BibitemOpen
  \bibfield  {author} {\bibinfo {author} {\bibfnamefont {Y.}~\bibnamefont
  {{Gu}}}\ and\ \bibinfo {author} {\bibfnamefont {J.}~\bibnamefont {{Wang}}},\
  }\bibfield  {title} {\bibinfo {title} {{Time evolution of coarse-grained
  entropy in classical and quantum motions of strongly chaotic systems}},\
  }\href {https://doi.org/10.1016/S0375-9601(97)00194-1} {\bibfield  {journal}
  {\bibinfo  {journal} {Phys. Lett. A}\ }\textbf {\bibinfo {volume} {229}},\
  \bibinfo {pages} {208} (\bibinfo {year} {1997})}\BibitemShut {NoStop}%
\bibitem [{\citenamefont {{Falcioni}}\ \emph {et~al.}(2005)\citenamefont
  {{Falcioni}}, \citenamefont {{Palatella}},\ and\ \citenamefont
  {{Vulpiani}}}]{falcioni2005production}%
  \BibitemOpen
  \bibfield  {author} {\bibinfo {author} {\bibfnamefont {M.}~\bibnamefont
  {{Falcioni}}}, \bibinfo {author} {\bibfnamefont {L.}~\bibnamefont
  {{Palatella}}},\ and\ \bibinfo {author} {\bibfnamefont {A.}~\bibnamefont
  {{Vulpiani}}},\ }\bibfield  {title} {\bibinfo {title} {{Production rate of
  the coarse-grained Gibbs entropy and the Kolmogorov-Sinai entropy: A real
  connection?}},\ }\href {https://doi.org/10.1103/PhysRevE.71.016118}
  {\bibfield  {journal} {\bibinfo  {journal} {\pre}\ }\textbf {\bibinfo
  {volume} {71}},\ \bibinfo {eid} {016118} (\bibinfo {year} {2005})},\ \Eprint
  {https://arxiv.org/abs/nlin/0407056} {arXiv:nlin/0407056 [nlin.CD]}
  \BibitemShut {NoStop}%
\bibitem [{\citenamefont {{Gell-Mann}}\ and\ \citenamefont
  {{Hartle}}(2007)}]{gellmann2007quasiclassical}%
  \BibitemOpen
  \bibfield  {author} {\bibinfo {author} {\bibfnamefont {M.}~\bibnamefont
  {{Gell-Mann}}}\ and\ \bibinfo {author} {\bibfnamefont {J.~B.}\ \bibnamefont
  {{Hartle}}},\ }\bibfield  {title} {\bibinfo {title} {{Quasiclassical coarse
  graining and thermodynamic entropy}},\ }\href
  {https://doi.org/10.1103/PhysRevA.76.022104} {\bibfield  {journal} {\bibinfo
  {journal} {\pra}\ }\textbf {\bibinfo {volume} {76}},\ \bibinfo {eid} {022104}
  (\bibinfo {year} {2007})},\ \Eprint {https://arxiv.org/abs/quant-ph/0609190}
  {arXiv:quant-ph/0609190 [quant-ph]} \BibitemShut {NoStop}%
\bibitem [{\citenamefont {{Kozlov}}\ and\ \citenamefont
  {{Treshchev}}(2007)}]{kozlov2007fine}%
  \BibitemOpen
  \bibfield  {author} {\bibinfo {author} {\bibfnamefont {V.~V.}\ \bibnamefont
  {{Kozlov}}}\ and\ \bibinfo {author} {\bibfnamefont {D.~V.}\ \bibnamefont
  {{Treshchev}}},\ }\bibfield  {title} {\bibinfo {title} {{Fine-grained and
  coarse-grained entropy in problems of statistical mechanics}},\ }\href
  {https://doi.org/10.1007/s11232-007-0040-1} {\bibfield  {journal} {\bibinfo
  {journal} {Theor. Math. Phys.}\ }\textbf {\bibinfo {volume} {151}},\ \bibinfo
  {pages} {539} (\bibinfo {year} {2007})}\BibitemShut {NoStop}%
\bibitem [{\citenamefont {{Rahav}}\ and\ \citenamefont
  {{Jarzynski}}(2007)}]{saar2007fluctuation}%
  \BibitemOpen
  \bibfield  {author} {\bibinfo {author} {\bibfnamefont {S.}~\bibnamefont
  {{Rahav}}}\ and\ \bibinfo {author} {\bibfnamefont {C.}~\bibnamefont
  {{Jarzynski}}},\ }\bibfield  {title} {\bibinfo {title} {{Fluctuation
  relations and coarse-graining}},\ }\href
  {https://doi.org/10.1088/1742-5468/2007/09/P09012} {\bibfield  {journal}
  {\bibinfo  {journal} {JSTAT}\ }\textbf {\bibinfo {volume} {2007}}\bibfield
  {number} {\bibinfo  {number} { (9)},\ \bibinfo {pages} {09012}},\ }\Eprint
  {https://arxiv.org/abs/0708.2437} {arXiv:0708.2437 [cond-mat.stat-mech]}
  \BibitemShut {NoStop}%
\bibitem [{\citenamefont {{Shell}}(2008)}]{shell2008relative}%
  \BibitemOpen
  \bibfield  {author} {\bibinfo {author} {\bibfnamefont {M.~S.}\ \bibnamefont
  {{Shell}}},\ }\bibfield  {title} {\bibinfo {title} {{The relative entropy is
  fundamental to multiscale and inverse thermodynamic problems}},\ }\href
  {https://doi.org/10.1063/1.2992060} {\bibfield  {journal} {\bibinfo
  {journal} {\jcp}\ }\textbf {\bibinfo {volume} {129}},\ \bibinfo {pages}
  {144108} (\bibinfo {year} {2008})}\BibitemShut {NoStop}%
\bibitem [{\citenamefont {{Puglisi}}\ \emph {et~al.}(2010)\citenamefont
  {{Puglisi}}, \citenamefont {{Pigolotti}}, \citenamefont {{Rondoni}},\ and\
  \citenamefont {{Vulpiani}}}]{puglisi2010entropy}%
  \BibitemOpen
  \bibfield  {author} {\bibinfo {author} {\bibfnamefont {A.}~\bibnamefont
  {{Puglisi}}}, \bibinfo {author} {\bibfnamefont {S.}~\bibnamefont
  {{Pigolotti}}}, \bibinfo {author} {\bibfnamefont {L.}~\bibnamefont
  {{Rondoni}}},\ and\ \bibinfo {author} {\bibfnamefont {A.}~\bibnamefont
  {{Vulpiani}}},\ }\bibfield  {title} {\bibinfo {title} {{Entropy production
  and coarse graining in Markov processes}},\ }\href
  {https://doi.org/10.1088/1742-5468/2010/05/P05015} {\bibfield  {journal}
  {\bibinfo  {journal} {JSTAT}\ }\textbf {\bibinfo {volume} {2010}}\bibfield
  {number} {\bibinfo  {number} { (5)},\ \bibinfo {pages} {05015}},\ }\Eprint
  {https://arxiv.org/abs/1002.4520} {arXiv:1002.4520 [cond-mat.stat-mech]}
  \BibitemShut {NoStop}%
\bibitem [{\citenamefont {{Gemmer}}\ and\ \citenamefont
  {{Steinigeweg}}(2014)}]{gemmer2014entropy}%
  \BibitemOpen
  \bibfield  {author} {\bibinfo {author} {\bibfnamefont {J.}~\bibnamefont
  {{Gemmer}}}\ and\ \bibinfo {author} {\bibfnamefont {R.}~\bibnamefont
  {{Steinigeweg}}},\ }\bibfield  {title} {\bibinfo {title} {{Entropy increase
  in K-step Markovian and consistent dynamics of closed quantum systems}},\
  }\href {https://doi.org/10.1103/PhysRevE.89.042113} {\bibfield  {journal}
  {\bibinfo  {journal} {\pre}\ }\textbf {\bibinfo {volume} {89}},\ \bibinfo
  {eid} {042113} (\bibinfo {year} {2014})}\BibitemShut {NoStop}%
\bibitem [{\citenamefont {Kelly}\ and\ \citenamefont
  {Wall}(2014)}]{kelly2014coarse}%
  \BibitemOpen
  \bibfield  {author} {\bibinfo {author} {\bibfnamefont {W.~R.}\ \bibnamefont
  {Kelly}}\ and\ \bibinfo {author} {\bibfnamefont {A.~C.}\ \bibnamefont
  {Wall}},\ }\bibfield  {title} {\bibinfo {title} {{Coarse-grained entropy and
  causal holographic information in AdS/CFT}},\ }\href
  {https://doi.org/10.1007/JHEP03(2014)118} {\bibfield  {journal} {\bibinfo
  {journal} {JHEP}\ }\textbf {\bibinfo {volume} {2014}},\ \bibinfo {pages}
  {118}},\ \Eprint {https://arxiv.org/abs/1309.3610} {arXiv:1309.3610 [hep-th]}
  \BibitemShut {NoStop}%
\bibitem [{\citenamefont {{Alonso-Serrano}}\ and\ \citenamefont
  {{Visser}}(2017)}]{alonso2017coarse}%
  \BibitemOpen
  \bibfield  {author} {\bibinfo {author} {\bibfnamefont {A.}~\bibnamefont
  {{Alonso-Serrano}}}\ and\ \bibinfo {author} {\bibfnamefont {M.}~\bibnamefont
  {{Visser}}},\ }\bibfield  {title} {\bibinfo {title} {{Coarse Graining Shannon
  and von Neumann Entropies}},\ }\href {https://doi.org/10.3390/e19050207}
  {\bibfield  {journal} {\bibinfo  {journal} {Entropy}\ }\textbf {\bibinfo
  {volume} {19}},\ \bibinfo {pages} {207} (\bibinfo {year} {2017})},\ \Eprint
  {https://arxiv.org/abs/1704.00237} {arXiv:1704.00237 [quant-ph]} \BibitemShut
  {NoStop}%
\bibitem [{\citenamefont {{Engelhardt}}\ and\ \citenamefont
  {{Wall}}(2018)}]{engelhardt2018decoding}%
  \BibitemOpen
  \bibfield  {author} {\bibinfo {author} {\bibfnamefont {N.}~\bibnamefont
  {{Engelhardt}}}\ and\ \bibinfo {author} {\bibfnamefont {A.~C.}\ \bibnamefont
  {{Wall}}},\ }\bibfield  {title} {\bibinfo {title} {{Decoding the Apparent
  Horizon: Coarse-Grained Holographic Entropy}},\ }\href
  {https://doi.org/10.1103/PhysRevLett.121.211301} {\bibfield  {journal}
  {\bibinfo  {journal} {\prl}\ }\textbf {\bibinfo {volume} {121}},\ \bibinfo
  {eid} {211301} (\bibinfo {year} {2018})},\ \Eprint
  {https://arxiv.org/abs/1706.02038} {arXiv:1706.02038 [hep-th]} \BibitemShut
  {NoStop}%
\bibitem [{\citenamefont {{Busiello}}\ \emph {et~al.}(2019)\citenamefont
  {{Busiello}}, \citenamefont {{Hidalgo}},\ and\ \citenamefont
  {{Maritan}}}]{busiello2019entropy}%
  \BibitemOpen
  \bibfield  {author} {\bibinfo {author} {\bibfnamefont {D.~M.}\ \bibnamefont
  {{Busiello}}}, \bibinfo {author} {\bibfnamefont {J.}~\bibnamefont
  {{Hidalgo}}},\ and\ \bibinfo {author} {\bibfnamefont {A.}~\bibnamefont
  {{Maritan}}},\ }\bibfield  {title} {\bibinfo {title} {{Entropy production for
  coarse-grained dynamics}},\ }\href {https://doi.org/10.1088/1367-2630/ab29c0}
  {\bibfield  {journal} {\bibinfo  {journal} {New J. Phys.}\ }\textbf {\bibinfo
  {volume} {21}},\ \bibinfo {eid} {073004} (\bibinfo {year} {2019})},\ \Eprint
  {https://arxiv.org/abs/1810.01833} {arXiv:1810.01833 [cond-mat.stat-mech]}
  \BibitemShut {NoStop}%
\bibitem [{\citenamefont {{Strasberg}}\ and\ \citenamefont
  {{Winter}}(2020{\natexlab{a}})}]{strasberg2020heat}%
  \BibitemOpen
  \bibfield  {author} {\bibinfo {author} {\bibfnamefont {P.}~\bibnamefont
  {{Strasberg}}}\ and\ \bibinfo {author} {\bibfnamefont {A.}~\bibnamefont
  {{Winter}}},\ }\bibfield  {title} {\bibinfo {title} {{Heat, Work and Entropy
  Production in Open Quantum Systems: A Microscopic Approach Based on
  Observational Entropy}},\ }\href@noop {} {\bibfield  {journal} {\bibinfo
  {journal} {eprint}\ } (\bibinfo {year} {2020}{\natexlab{a}})},\ \Eprint
  {https://arxiv.org/abs/2002.08817v1} {arXiv:2002.08817v1 [quant-ph]}
  \BibitemShut {NoStop}%
\bibitem [{\citenamefont {{{\v{S}}afr{\'a}nek}}\ \emph
  {et~al.}(2019{\natexlab{a}})\citenamefont {{{\v{S}}afr{\'a}nek}},
  \citenamefont {{Deutsch}},\ and\ \citenamefont {{Aguirre}}}]{Safranek2019a}%
  \BibitemOpen
  \bibfield  {author} {\bibinfo {author} {\bibfnamefont {D.}~\bibnamefont
  {{{\v{S}}afr{\'a}nek}}}, \bibinfo {author} {\bibfnamefont {J.~M.}\
  \bibnamefont {{Deutsch}}},\ and\ \bibinfo {author} {\bibfnamefont
  {A.}~\bibnamefont {{Aguirre}}},\ }\bibfield  {title} {\bibinfo {title}
  {{Quantum coarse-grained entropy and thermodynamics}},\ }\href
  {https://doi.org/10.1103/PhysRevA.99.010101} {\bibfield  {journal} {\bibinfo
  {journal} {\pra}\ }\textbf {\bibinfo {volume} {99}},\ \bibinfo {eid} {010101}
  (\bibinfo {year} {2019}{\natexlab{a}})},\ \Eprint
  {https://arxiv.org/abs/1707.09722} {arXiv:1707.09722 [quant-ph]} \BibitemShut
  {NoStop}%
\bibitem [{\citenamefont {{{\v{S}}afr{\'a}nek}}\ \emph
  {et~al.}(2019{\natexlab{b}})\citenamefont {{{\v{S}}afr{\'a}nek}},
  \citenamefont {{Deutsch}},\ and\ \citenamefont {{Aguirre}}}]{Safranek2019b}%
  \BibitemOpen
  \bibfield  {author} {\bibinfo {author} {\bibfnamefont {D.}~\bibnamefont
  {{{\v{S}}afr{\'a}nek}}}, \bibinfo {author} {\bibfnamefont {J.~M.}\
  \bibnamefont {{Deutsch}}},\ and\ \bibinfo {author} {\bibfnamefont
  {A.}~\bibnamefont {{Aguirre}}},\ }\bibfield  {title} {\bibinfo {title}
  {{Quantum coarse-grained entropy and thermalization in closed systems}},\
  }\href {https://doi.org/10.1103/PhysRevA.99.012103} {\bibfield  {journal}
  {\bibinfo  {journal} {\pra}\ }\textbf {\bibinfo {volume} {99}},\ \bibinfo
  {eid} {012103} (\bibinfo {year} {2019}{\natexlab{b}})},\ \Eprint
  {https://arxiv.org/abs/1803.00665} {arXiv:1803.00665 [quant-ph]} \BibitemShut
  {NoStop}%
\bibitem [{\citenamefont {{{\v{S}}afr{\'a}nek}}\ \emph
  {et~al.}(2019{\natexlab{c}})\citenamefont {{{\v{S}}afr{\'a}nek}},
  \citenamefont {{Aguirre}},\ and\ \citenamefont
  {{Deutsch}}}]{safranek2019classical}%
  \BibitemOpen
  \bibfield  {author} {\bibinfo {author} {\bibfnamefont {D.}~\bibnamefont
  {{{\v{S}}afr{\'a}nek}}}, \bibinfo {author} {\bibfnamefont {A.}~\bibnamefont
  {{Aguirre}}},\ and\ \bibinfo {author} {\bibfnamefont {J.~M.}\ \bibnamefont
  {{Deutsch}}},\ }\bibfield  {title} {\bibinfo {title} {{Classical dynamical
  coarse-grained entropy and comparison with the quantum version}},\
  }\href@noop {} {\bibfield  {journal} {\bibinfo  {journal} {eprint}\ }
  (\bibinfo {year} {2019}{\natexlab{c}})},\ \Eprint
  {https://arxiv.org/abs/1905.03841} {arXiv:1905.03841 [cond-mat.stat-mech]}
  \BibitemShut {NoStop}%
\bibitem [{\citenamefont {{Strasberg}}\ and\ \citenamefont
  {{Winter}}(2020{\natexlab{b}})}]{strasberg2020dissipation}%
  \BibitemOpen
  \bibfield  {author} {\bibinfo {author} {\bibfnamefont {P.}~\bibnamefont
  {{Strasberg}}}\ and\ \bibinfo {author} {\bibfnamefont {A.}~\bibnamefont
  {{Winter}}},\ }\bibfield  {title} {\bibinfo {title} {{Dissipation in Quantum
  Systems: A Unifying Picture}},\ }\href@noop {} {\bibfield  {journal}
  {\bibinfo  {journal} {eprint}\ } (\bibinfo {year} {2020}{\natexlab{b}})},\
  \Eprint {https://arxiv.org/abs/2002.08817v2} {arXiv:2002.08817v2 [quant-ph]}
  \BibitemShut {NoStop}%
\bibitem [{\citenamefont {{Strasberg}}(2019)}]{strasberg2019entropy}%
  \BibitemOpen
  \bibfield  {author} {\bibinfo {author} {\bibfnamefont {P.}~\bibnamefont
  {{Strasberg}}},\ }\bibfield  {title} {\bibinfo {title} {{Entropy production
  as change in observational entropy}},\ }\href@noop {} {\bibfield  {journal}
  {\bibinfo  {journal} {eprint}\ } (\bibinfo {year} {2019})},\ \Eprint
  {https://arxiv.org/abs/1906.09933} {arXiv:1906.09933 [cond-mat.stat-mech]}
  \BibitemShut {NoStop}%
\bibitem [{\citenamefont {{Schindler}}\ \emph {et~al.}(2020)\citenamefont
  {{Schindler}}, \citenamefont {{{\v{S}}afr{\'a}nek}},\ and\ \citenamefont
  {{Aguirre}}}]{schindler2020entanglement}%
  \BibitemOpen
  \bibfield  {author} {\bibinfo {author} {\bibfnamefont {J.}~\bibnamefont
  {{Schindler}}}, \bibinfo {author} {\bibfnamefont {D.}~\bibnamefont
  {{{\v{S}}afr{\'a}nek}}},\ and\ \bibinfo {author} {\bibfnamefont
  {A.}~\bibnamefont {{Aguirre}}},\ }\bibfield  {title} {\bibinfo {title}
  {{Entanglement entropy from coarse-graining in pure and mixed multipartite
  systems}},\ }\href@noop {} {\bibfield  {journal} {\bibinfo  {journal}
  {eprint}\ } (\bibinfo {year} {2020})},\ \Eprint
  {https://arxiv.org/abs/2005.05408} {arXiv:2005.05408 [quant-ph]} \BibitemShut
  {NoStop}%
\bibitem [{\citenamefont {{Modi}}\ \emph {et~al.}(2010)\citenamefont {{Modi}},
  \citenamefont {{Paterek}}, \citenamefont {{Son}}, \citenamefont {{Vedral}},\
  and\ \citenamefont {{Williamson}}}]{modi2010unified}%
  \BibitemOpen
  \bibfield  {author} {\bibinfo {author} {\bibfnamefont {K.}~\bibnamefont
  {{Modi}}}, \bibinfo {author} {\bibfnamefont {T.}~\bibnamefont {{Paterek}}},
  \bibinfo {author} {\bibfnamefont {W.}~\bibnamefont {{Son}}}, \bibinfo
  {author} {\bibfnamefont {V.}~\bibnamefont {{Vedral}}},\ and\ \bibinfo
  {author} {\bibfnamefont {M.}~\bibnamefont {{Williamson}}},\ }\bibfield
  {title} {\bibinfo {title} {{Unified View of Quantum and Classical
  Correlations}},\ }\href {https://doi.org/10.1103/PhysRevLett.104.080501}
  {\bibfield  {journal} {\bibinfo  {journal} {\prl}\ }\textbf {\bibinfo
  {volume} {104}},\ \bibinfo {eid} {080501} (\bibinfo {year} {2010})},\ \Eprint
  {https://arxiv.org/abs/0911.5417} {arXiv:0911.5417 [quant-ph]} \BibitemShut
  {NoStop}%
\bibitem [{\citenamefont {{Groisman}}\ \emph {et~al.}(2007)\citenamefont
  {{Groisman}}, \citenamefont {{Kenigsberg}},\ and\ \citenamefont
  {{Mor}}}]{groisman2007quantumness}%
  \BibitemOpen
  \bibfield  {author} {\bibinfo {author} {\bibfnamefont {B.}~\bibnamefont
  {{Groisman}}}, \bibinfo {author} {\bibfnamefont {D.}~\bibnamefont
  {{Kenigsberg}}},\ and\ \bibinfo {author} {\bibfnamefont {T.}~\bibnamefont
  {{Mor}}},\ }\bibfield  {title} {\bibinfo {title} {{``Quantumness'' versus
  ``Classicality'' of Quantum States}},\ }\href@noop {} {\bibfield  {journal}
  {\bibinfo  {journal} {arXiv e-prints}\ ,\ \bibinfo {eid} {quant-ph/0703103}}
  (\bibinfo {year} {2007})},\ \Eprint {https://arxiv.org/abs/quant-ph/0703103}
  {arXiv:quant-ph/0703103 [quant-ph]} \BibitemShut {NoStop}%
\bibitem [{\citenamefont {{Piani}}\ \emph {et~al.}(2011)\citenamefont
  {{Piani}}, \citenamefont {{Gharibian}}, \citenamefont {{Adesso}},
  \citenamefont {{Calsamiglia}}, \citenamefont {{Horodecki}},\ and\
  \citenamefont {{Winter}}}]{piani2011all}%
  \BibitemOpen
  \bibfield  {author} {\bibinfo {author} {\bibfnamefont {M.}~\bibnamefont
  {{Piani}}}, \bibinfo {author} {\bibfnamefont {S.}~\bibnamefont
  {{Gharibian}}}, \bibinfo {author} {\bibfnamefont {G.}~\bibnamefont
  {{Adesso}}}, \bibinfo {author} {\bibfnamefont {J.}~\bibnamefont
  {{Calsamiglia}}}, \bibinfo {author} {\bibfnamefont {P.}~\bibnamefont
  {{Horodecki}}},\ and\ \bibinfo {author} {\bibfnamefont {A.}~\bibnamefont
  {{Winter}}},\ }\bibfield  {title} {\bibinfo {title} {{All Nonclassical
  Correlations Can Be Activated into Distillable Entanglement}},\ }\href
  {https://doi.org/10.1103/PhysRevLett.106.220403} {\bibfield  {journal}
  {\bibinfo  {journal} {\prl}\ }\textbf {\bibinfo {volume} {106}},\ \bibinfo
  {eid} {220403} (\bibinfo {year} {2011})},\ \Eprint
  {https://arxiv.org/abs/1103.4032} {arXiv:1103.4032 [quant-ph]} \BibitemShut
  {NoStop}%
\bibitem [{\citenamefont {{Horodecki}}\ \emph {et~al.}(2005)\citenamefont
  {{Horodecki}}, \citenamefont {{Horodecki}}, \citenamefont {{Horodecki}},
  \citenamefont {{Oppenheim}}, \citenamefont {{Sen(de)}}, \citenamefont
  {{Sen}},\ and\ \citenamefont {{Synak-Radtke}}}]{horodecki2005local}%
  \BibitemOpen
  \bibfield  {author} {\bibinfo {author} {\bibfnamefont {M.}~\bibnamefont
  {{Horodecki}}}, \bibinfo {author} {\bibfnamefont {P.}~\bibnamefont
  {{Horodecki}}}, \bibinfo {author} {\bibfnamefont {R.}~\bibnamefont
  {{Horodecki}}}, \bibinfo {author} {\bibfnamefont {J.}~\bibnamefont
  {{Oppenheim}}}, \bibinfo {author} {\bibfnamefont {A.}~\bibnamefont
  {{Sen(de)}}}, \bibinfo {author} {\bibfnamefont {U.}~\bibnamefont {{Sen}}},\
  and\ \bibinfo {author} {\bibfnamefont {B.}~\bibnamefont {{Synak-Radtke}}},\
  }\bibfield  {title} {\bibinfo {title} {{Local versus nonlocal information in
  quantum-information theory: Formalism and phenomena}},\ }\href
  {https://doi.org/10.1103/PhysRevA.71.062307} {\bibfield  {journal} {\bibinfo
  {journal} {\pra}\ }\textbf {\bibinfo {volume} {71}},\ \bibinfo {eid} {062307}
  (\bibinfo {year} {2005})},\ \Eprint {https://arxiv.org/abs/quant-ph/0410090}
  {arXiv:quant-ph/0410090 [quant-ph]} \BibitemShut {NoStop}%
\bibitem [{\citenamefont
  {{{\v{S}}afr{\'a}nek}}(2020)}]{safranek2020observational}%
  \BibitemOpen
  \bibfield  {author} {\bibinfo {author} {\bibfnamefont {D.}~\bibnamefont
  {{{\v{S}}afr{\'a}nek}}},\ }\bibfield  {title} {\bibinfo {title}
  {{Observational entropy with generalized measurements}},\ }\href@noop {}
  {\bibfield  {journal} {\bibinfo  {journal} {arXiv e-prints}\ ,\ \bibinfo
  {eid} {arXiv:2007.07246}} (\bibinfo {year} {2020})},\ \Eprint
  {https://arxiv.org/abs/2007.07246} {arXiv:2007.07246 [quant-ph]} \BibitemShut
  {NoStop}%
\end{thebibliography}%


\end{document}